\documentclass[12pt]{extarticle}
\usepackage{geometry}
\usepackage{todonotes} 
\usepackage{subfig}
\makeatletter
\newcommand\ackname{Acknowledgements}
\if@titlepage
\newenvironment{acknowledgements}{
	\titlepage
	\null\vfil
	\@beginparpenalty\@lowpenalty
	\begin{center}
		\bfseries \ackname
		\@endparpenalty\@M1
\end{center}}
{\par\vfil\null\endtitlepage}
\else

\usepackage{mathrsfs}
\usepackage{eucal}
\usepackage{amssymb,amsmath}

\usepackage{pgf}
\usepackage{tikz}
\usepackage{tikz-3dplot}
\usetikzlibrary{decorations.pathreplacing,calligraphy,positioning}
\usepackage{cite}
\usepackage{graphicx}
\usepackage{caption}
\usepackage{color}
\usepackage{amssymb}
\usepgflibrary{arrows}
\usetikzlibrary{calc,angles,positioning,intersections,quotes,backgrounds,decorations,decorations.markings,decorations.text,decorations.pathreplacing}
\usepackage{amsmath}
\usepackage{amsthm}
\usepackage{accents}

\theoremstyle{remark}
\theoremstyle{definition}

\usepackage{thmtools, thm-restate}

\numberwithin{equation}{section}

\usepackage{comment}
\usepackage{physics}
\definecolor{mediumtaupe}{rgb}{0.4, 0.3, 0.28}
\definecolor{skobeloff}{rgb}{0.0, 0.48, 0.45}
\definecolor{sandstorm}{rgb}{0.93, 0.84, 0.25}
\definecolor{khaki}{rgb}{0.76, 0.69, 0.57}
\definecolor{olivedrab7}{rgb}{0.24, 0.2, 0.12}
\definecolor{sanddune}{rgb}{0.59, 0.44, 0.09}
\definecolor{mediumseagreen}{rgb}{0.24, 0.7, 0.44}
\definecolor{persianplum}{rgb}{0.44, 0.11, 0.11}

\DeclareSymbolFont{rsfs}{U}{rsfs}{m}{n}
\DeclareSymbolFontAlphabet{\mathscrsfs}{rsfs}
\makeatletter
\newcommand*{\rom}[1]{\expandafter\@slowromancap\romannumeral #1@}
\makeatother

\usetikzlibrary{decorations.pathreplacing,decorations.markings}
\tikzset{
	on each segment/.style={
		decorate,
		decoration={
			show path construction,
			moveto code={},
			lineto code={
				\path [#1]
				(\tikzinputsegmentfirst) -- (\tikzinputsegmentlast);
			},
			curveto code={
				\path [#1] (\tikzinputsegmentfirst)
				.. controls
				(\tikzinputsegmentsupporta) and (\tikzinputsegmentsupportb)
				..
				(\tikzinputsegmentlast);
			},
			closepath code={
				\path [#1]
				(\tikzinputsegmentfirst) -- (\tikzinputsegmentlast);
			},
		},
	},
	mid arrow/.style={postaction={decorate,decoration={
				markings,
				mark=at position .5 with {\arrow[#1]{stealth}}
	}}},
}
\begin{document}
	\title{\textbf{Lagrangian 1-form structure of Calogero-Moser type systems}}
	\author{Thanadon Kongkoom$^{1,*}$, Frank W. Nijhoff$^{2,\dagger}$ and  Sikarin Yoo-Kong$^{1,\ddagger}$ \\
            \small {$^1$The Institute for Fundamental Study (IF),} \small\emph{Naresuan University, Phitsanulok, Thailand, 65000.}\\
		\small {$^2$School of Mathematics, University of Leeds ,} \small\emph{ Leeds LS2 9JT, United Kingdom.}\\
		\small{$^*$thanadonko63@nu.ac.th, $^\dagger$frank.nijhoff@gmail.com, $^\ddagger$sikariny@nu.ac.th} \\
	}
	\date{}
	\maketitle
	\abstract
   
    We consider the variational principle for the Lagrangian 1-form structure for long-range 
    models of Calogero-Moser (CM) type. The multiform variational principle involves variations 
    with respect to both the field variables as well as the independent variables 
    corresponding to deformations of the time-curves in a multi-time space. The ensuing 
    generalised Euler-Lagrange (gEL) equations comprise a system of multi-time 
    EL equations, as well as constraints from so-called `alien derivatives' and `corner 
    equations' arising from how variations on different coordinate curves match up. 
    The closure relation, i.e. closedness of the Lagrange 1-form on solutions of the EL 
    system, guarantees the stationarity of the action functional under deformation of the 
    time-curves, and hence the multidimensional consistency of the corresponding gEL 
    system. Using this as an integrability criterion on the Lagrangian level, we apply the 
    system to some ans\"atze on the kinetic form of the Lagrangian components, associated 
    with models of CM type without specifying the potentials. We show that from this 
    integrability criterion the general elliptic form of the three systems, 
    Calogero-Moser, Ruijsenaars-Schneider, and Goldfish systems, can be derived. 
    We extend the analysis to an associated Hamiltonian formalism, via Noether's theorem 
    and by applying Legendre transformations. Thus, the multiform variational principle 
    leads to a system of generalised Hamilton equations describing Hamiltonian commuting flows for the mentioned elliptic models. 
	\\ 
 \\
	\textbf{Keywords}: Integrability, Elliptic potential, Multidimensional consistency, Variational principle, Lagrangian 1-forms, Integrable Lagrangian dynamics, Integrable Hamiltonian dynamics, Calogero-Moser type systems
	\section{Introduction}
    In the past 15 years, it has become evident that the proper variational description of 
   integrable systems in the sense of \textit{multidimensional consistency} (MDC) is that of Lagrangian multiform theory, which was initiated in \cite{S.Lobb}. 
    The key idea behind this this novel variational formalism is to account for the 
    phenomenon of multidimensional consistency (MDC) where the integrability manifests 
    itself by the existence of a multitude of compatible equations on one and the same 
    dependent variable (which could be a multi-component variable) with regard to a 
    dependence on a higher-dimensional multi-time space of independent variables. 
    While the first examples of multiforms comprised discrete 2-form and 3-form cases 
    \cite{S.Lobb, Lobb_2009}, the study of 1-forms, $\mathscr{L}=\sum_i L_i\mathrm{d}t_i$ 
    in a multi-time space of variables $\boldsymbol{t}=(t_1,t_2,\cdots)$, with coefficients 
    $L_i$, was essential to gain an understanding of how the new variational principle 
    worked, and the Calogero-Moser (CM) system, both in continuous as well as discrete 
    multi-time, was used as a laboratory, cf. \cite{RinCM, Rinthesis, RinRS}. 
    The CM model has the advantage that (both in the discrete- as well as continuous time 
    case) the general solution can be 
    given in terms of a secular problem. Furthermore, the discrete-time model contains parameters which are inherited from 
    a corresponding Kadomtsev-Petviashvili system, from which it can be derived as a pole reduction, cf. also \cite{Krich}. 
    The variational principle for the Lagrangian 1-form $\mathscr{L}$ was shown to give rise to a system of 
    generalised Euler-Lagrange (gEL) equations, comprising the following elements: 
    \begin{itemize}
    \item Euler-Lagrange equations along each of the coordinate axes; 
    \item corner equation arising from stepped curves (where coordinate lines meet); 
    \item constraints arising from `alien derivatives (derivatives w.r.t. variables not belonging to the 
    coordinate lines); 
    \item closure relations  $\partial L_i/\partial t_j=\partial L_j/\partial t_i,\;\forall i\neq j$ guaranteeing 
    the invariance of the action under deformations of the coordinate curves. 
    \end{itemize} 
    It is conjectured that any system of Lagrangians given by the 1-form satisfying this set of equations will be integrable in the sense of multidimensional consistency (i.e. in the sense of commuting flows). However, one could change the perspective 
    and consider the set of gEL equations as the defining set for Lagrangian 1-forms, predicting the form of the Lagrangians 
    themselves. Therefore, it is possible to use this set of equations to search for or classify integrable systems. 
    Since the initial work opening up the subject, the investigation of Lagrangian multiforms was advanced over the past 
    decade, \cite{Xenitidis, Suris1, Jairuk2, Jairuk3, Petrera, Suris2, Mats2, Mats3}. Moreover, the Lagrangian 1-form was developed to study in other models, e.g., Toda-type system \cite{10.1093/integr/xyy020, Petrera, Suris2, Suris_2018}, and the special case of non-parametrised formula of Ruijsenaar-Schneider model, namely the Goldfish system \cite{Jairuk1, PIENSUK202145}.
    \\
    \\
     For finite degree of freedom Hamiltonian systems, where the equations of motion are ODEs, it is the notion of Arnol'd-Liouville integrability (ODE system)\cite{Arnoldtextbook} that is usually adopted. This can also be extended to PDEs as Hamiltonian systems with an infinite degree of freedom. In this approach, for a system with a finite degree of freedom, there exists a collection of invariant quantities, which we can treat as a collection of Hamiltonians $(H_1,H_2,...,H_N)$, which are in involution: $ \{H_i,H_j\} = 0\;,\;\forall i\neq j$, with regard to a 
     appropriate choice of Poisson structure, leading to a key feature known as the Hamiltonian commuting flows. The connection between the Lagrangian 1-form and Hamiltonian system was given by Suris \cite{Suris1}, see also \cite{Puttarprom2019IntegrableHH}. 
     Like in the Lagrangian multiform approach, the Hamiltonian multiform principle leads to 
     a set of generalised Hamilton equations, which can be considered to constitute an 
     integrable multi-time system.
    \\
    \\
    Since integrable systems are highly special, the search and classification for integrable systems has always been one of the challenging tasks of the field. However, since this kind of systems are equipped with intriguing mathematical structures, there are many tools 
    are readily available. Lagrangian multiform structures seem to be a fundamental and universal structure of integrability, and we can adopt its existence as a defining 
    property of integrability.  From this perspective, we start, in the present work, from an ansatz for 
    the form of the first two Lagrangian components of a Lagrangian 1-form (which is 
    sufficient for the 2-time case), but containing \textit{ab initio} arbitrary potential functions,  for the Calogero-Moser (CM), the Goldfish (GF), and the Ruijsenaar-Schneider (RS) type, and 
    derive from the multiform gEL equation the explicit form of the potential functions corresponding to the well-known integrable cases. Thus, we show that already the 
    2-time multiform structure is sufficient to select the integrable potentials. 
    \\
    \\
    The organisation of the paper is as follows. In section \ref{ssection2.1}, the integrability criterion for Lagrangian 1-forms, i.e., the system of closure relations and constraints with the condition \textit{on-shell} will be briefly given. Then, in section \ref{ssection2.2}, we show how to derive the elliptic potentials for models of CM, GF, and RS  
    from the 1-form principle. In section \ref{ssection3.1}, the connection between Liouville integrability and the 
    Hamiltonian 1-form formalism, associated with a system of Hamiltonian commuting flows with the condition \textit{on-shell}, will be established. In section \ref{ssection3.2}, the elliptic potentials, which are identical with the ones obtained from the Lagrangian approach for CM, GF, and RS models,  are obtained. Finally, the conclusion and suggestions for further investigation are provided in section \ref{section4}.
 \section{Lagrangian 1-form structure}\label{section2}

     \subsection{A brief review of the Lagrangian 1-form theory }\label{ssection2.1}
     
     In this section, we will briefly describe the variational principle
     for a system of $N$-particles described by a set of position 
     variables $\mathbf{X} = (X_1,...,X_N)$ as functions $\mathbf{X}(\mathbf{t})$ as functions of 
     multi-time $\mathbf{t}=t_1,\dots,t_M$ (where $M$ in general does not need to be the same as $N$, but for the case of independent commuting time-flows they do coincide).  
     Furthermore, consider 
     parametrised curves $\Gamma$  in the $M$-dimensional space of time-variables 
     $t_i$, which (assuming them for simplicity to be smooth), which can be parametrised through a parameter $s\in[0,1]$, as $\mathbf{t}(s)=(t_1(s),...,t_M(s))$. As in finite-dimensional systems of Calogero-Moser type the vector fields corresponding to the higher time-flows $t_k$ for $k>N$ are not independent from those of the lower ones, we will assume henceforth that the dimension of the 
     multi-time spaces we consider will not exceed the number of degree of freedom, hence 
     we assume from now on that $M\leq N$, and for convenience we will set $M=N$ in most of the formulae below.

\begin{figure}[h]
    	    \centering
        	    \begin{tikzpicture}[scale = 0.8]
            	    \path[line width = 1pt, draw = black,->]
 	                  (0,0) to (5,0);
 	                  
 	                  \path[line width = 1pt, draw = black,->]
 	                  (0,0) to (0,3.5);

                        \path[line width = 1pt, draw = black,->]
 	                  (0,0) to (-2,-2);
 	                  
 	                \path[line width = 1pt, draw = black]
 	                  (0.5,-0.25) to [bend right = 50] (3,-1.5);

                        \path[dashed, line width = 1pt, draw = black]
 	                  (0.5,-0.25) to [bend left = 10] (3,-1.5);
 	                 
 	                \path[line width = 1pt, draw = black]
 	                  (0.5,2.75) to [bend right = 40] (3,1.5);

                        \path[dashed, line width = 1pt, draw = black]
 	                  (0.5,2.75) to [bend left = 20] (3,1.5);
                    
                        \path[dashed, line width = 0.5pt, draw = black]
 	                  (0.5,-0.25) to (0.5,2.75);
                        \path[dashed, line width = 0.5pt, draw = black]
 	                  (3,-1.5) to (3,1.5);
                    
 	                 \node at (5.25,0) {$t_i$};
 	                 \node at (-2,-1.5) {$t_j$};
 	                 \node at (0,3.75) {\scriptsize{$\mathbf{X}$}};

                        \node[circle, fill, inner sep=1.25 pt] at (0.5,-0.25) {};
                        \node[circle, fill, inner sep=1.25 pt] at (3,-1.5) {};
 	                \node at (0.15,-0.45) {\scriptsize{$\mathbf{t}(0)$}};
 	                \node at (3.5,-1.5) {\scriptsize{$\mathbf{t}(1)$}};

                        \node[circle, fill, inner sep=1.25 pt] at (0.5,2.75) {};
                        \node[circle, fill, inner sep=1.25 pt] at (3,1.5) {};
 	                \node at (1,3) {\scriptsize{$\mathbf{X}(\mathbf{t}(0))$}};
 	                \node at (3.75,1.5) {\scriptsize{$\mathbf{X}(\mathbf{t}(1))$}};
                   
 	                 \node at (2.5,-0.75) {\scriptsize{$\Gamma'$}};
 	                 \node at (0.75,-1.25) {\scriptsize{$\Gamma$}};
                        \node at (1.25,1.5) {\scriptsize{$\mathcal{E}_\Gamma$}};
 	                 \node at (2.5,2.25) {\scriptsize{$\mathcal{E}'_\Gamma$}};
 	                 
 	            \end{tikzpicture}
              \captionof{figure}{The curves $\Gamma$ and $\mathcal{E}_\Gamma$ in the $\mathbf{X}-\mathbf{t}$ configuration.}\label{fig1}
        \end{figure}
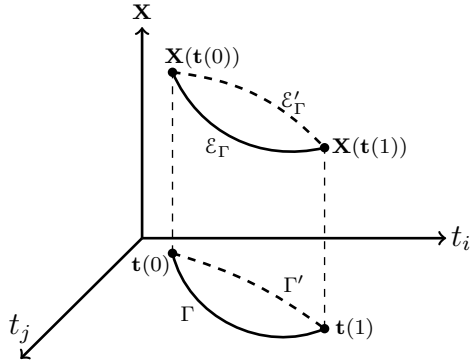

     The 1-form variational principle, cf. \cite{Frankbook}, can be thought of as consisting two processes, 
     illustrated in in Fig.\ref{fig1}: {\it i)} variations of the dependent variables $X_i(\mathbf{t})$ as functions of multi-time $\mathbf{t}$, and {\it ii)} variations of the curve $\Gamma$ in the 
     multi-time space, while fixing the end points. 
     Introducing Lagrangian components $L_i$ as functions of the dependent variables and their derivatives, e.g. in the first-jet case 
     $  L_i = L_i\left[\mathbf{X}(\mathbf{t}),\frac{\partial \mathbf{X}(\mathbf{t})}{\partial t_1},...,\frac{\partial \mathbf{X}(\mathbf{t})}{\partial t_N}\right]\;, $
     the multiform action is a functional of both the particle positions 
     $\mathbf{X}_i$ as well as of the curve $\Gamma$, and is given by
     \begin{equation}
         S[\mathbf{X}(\mathbf{t}),\Gamma] = \int_\Gamma\;\sum_{i=1}^NL_idt_i = \int_0^1\left(\sum_{i=1}^NL_i\frac{dt_i}{ds}\right)ds\;,\label{action}
     \end{equation}
     where in the last expression we have used the parametrisation of the curve $\Gamma$. 
    The two-step variational process described above is elaborated as follows. 
     We first consider the variation of the independent variables $\mathbf{t}(s)$ in the following way $\mathbf{t}(s)\to\mathbf{t}(s)+\delta\mathbf{t}(s)$, i.e., $\Gamma\to\Gamma'$, given in Fig.\ref{fig1}, keeping the end-points fixed, $\delta\mathbf{t}(0)=\delta\mathbf{t}(1)=0$. We consider $\delta S = S'[\mathbf{X}(\mathbf{t}+\delta \mathbf{t}),\Gamma']-S[\mathbf{X}(\mathbf{t}),\Gamma]$ and apply the Taylor expansion with the boundary conditions, resulting in
     \begin{align}
         \delta S =& \int_0^1ds\sum_{i=1}^N\left(\sum_{j=1}^N\frac{\partial L_i}{\partial t_j}\frac{dt_i}{ds}\delta t_j + L_i\frac{d(\delta t_i)}{ds}\right) \nonumber
         \\=& \int_0^1ds\sum_{i=1}^N\left(\sum_{j=1}^N\frac{\partial L_i}{\partial t_j}\frac{dt_i}{ds}\delta t_j - \frac{dL_i}{ds}\delta t_i\right) + \sum_{i=1}^N L_i\delta t_i(s)\bigg|_{s=0}^{s=1} \nonumber
         \\=& \int_0^1ds\sum_{i\neq j}^N\delta t_j\left(\frac{\partial L_i}{\partial t_j}-\frac{\partial L_j}{\partial t_i}\right)\frac{dt_i}{ds}\;.
     \end{align}
     Under the critical condition $\delta S=0$, the coefficients in integrand must be zero 
     leading to \textit{closure relation}:
     \begin{equation}
         \frac{\partial L_i}{\partial t_j}=\frac{\partial L_j}{\partial t_i}\;,\;\;\text{where}\;\;i,j=1,2,...,N\;,\label{closure}
     \end{equation}
     subject to the equations of motion. The closure relation guarantees that any such 
     continuous curve $\Gamma$ could be locally deformed without changing the value of the 
     action functional. Next, in order to obtain the EL equations, we now perform the variation 
     on the dependent variables $\mathbf{X}$. We now fix the curve $\Gamma$ on the independent 
     variable space and examine the corresponding evaluated curve $\mathcal{E}_\Gamma$ in the 
     extended space of independent variables $\mathbf{t}$ and field variables $\mathbf{X}$ given 
     in Fig.\ref{fig1}. The variation of critical action \eqref{action} subjects to the variation on configuration curve $\mathcal{E}
     _\Gamma\to\mathcal{E}'_\Gamma$, shown in Fig.\ref{fig1}, i.e., $\mathbf{X}\to\mathbf{X}+\delta\mathbf{X}$, 
      with the boundary conditions $
     \delta\mathbf{X}(\mathbf{t}(0))=\delta\mathbf{X}(\mathbf{t}(1))=0$, yielding
         \begin{subequations}\label{contEL}\begin{align}
         0 = \sum_{i=1}^N\frac{\partial L_i}{\partial \mathbf{X}}\frac{dt_i}{ds}-\frac{d}{ds}\left\{\frac{1}{\|d\mathbf{t}/ds\|^2}\sum_{i,j=1}^N\frac{\partial L_i}{\partial(\partial\mathbf{X}/\partial t_j)}\left(\frac{d t_i}{ds}\right)\left(\frac{d t_j}{ds}\right)\right\}\;,\label{dx}
         \end{align}
         and,
         \begin{align}
         0 =& \frac{\partial L_j}{\partial(\partial\mathbf{X}/\partial t_i)}\left(\frac{d t_j}{ds}\right)^2 - \left(\frac{\partial L_i}{\partial(\partial\mathbf{X}/\partial t_i)}-\frac{\partial L_j}{\partial(\partial\mathbf{X}/\partial t_j)}\right)\frac{d t_i}{ds}\frac{d t_j}{ds}  - \frac{\partial L_i}{\partial(\partial\mathbf{X}/\partial t_j)}\left(\frac{d t_i}{ds}\right)^2 \nonumber
         \\&+ \sum_{\substack{k=1\\k\neq i\neq j}}^N\left(\frac{\partial L_k}{\partial(\partial\mathbf{X}/\partial t_i)}\frac{d t_j}{ds}-\frac{\partial L_k}{\partial(\partial\mathbf{X}/\partial t_j)}\frac{d t_i}{ds}\right)\frac{dt_k}{ds}\;,\label{dy}
     \end{align}\end{subequations} 
     for all $j \neq i$ (for details of the derivation see appendix \ref{Appendix}). Since the closure relation \eqref{closure} implies that the action is invariant under the deformation of a smooth curve $\Gamma$, see \cite{RinCM}, \eqref{dx}-\eqref{dy} are the conditions for criticality on any given smooth curve $\Gamma$. However, it 
     is often computationally more convenient to consider the conditions on `stepped curves' which are piecewise made up from coordinate lines, cf. \cite{Suris_2016}, which leads to the following system of 
    \textit{generalised Euler-Lagrange equation} (gEL) equations, \cite{Suris2}:
 \begin{subequations}\label{eq:gEL}\label{gELCorner}
     \begin{align}
     {\textrm{\bf coordinate\ \ EL\ \ eqs.:}}  \qquad   &\frac{\partial L_i}{\partial\mathbf{X}}-\frac{\partial}{\partial t_i}\left(\frac{\partial L_i}{\partial(\partial\mathbf{X}/\partial t_i)}\right)=0\;,\label{gEL}\\ 
     {\textrm{\bf corner\ \ eqs.:}}\qquad \quad 
         &\frac{\partial L_i}{\partial \left(\partial\mathbf{X}/\partial t_i\right)}-\frac{\partial L_j}{\partial (\partial\mathbf{X}/\partial t_j)}=0\;,\label{corner2}\\
    {\textrm{\bf alien derivatives constraints:}} \quad     & \frac{\partial L_i}{\partial(\partial\mathbf{X}/\partial t_j)}=0\;,\quad \forall i\neq j\,, \label{corner}
     \end{align}
     \end{subequations}
     where $\forall i\neq j$ \cite{Suris2}. 
     We have called the various parts of the system \eqref{gELCorner} \textit{coordinate Euler-Lagrange equation} which are the usual EL 
     equations along each coordinate line, \textit{corner equation}\footnote{Corner equations first appeared in \cite{RinCM}, in the context of the discrete-time Calogero-Moser model, and  formed an integral 
     part of the relevant EL equations as in Chapter 3 of \cite{Rinthesis}. }, which take into account 
     variations (internal to the curve) at the points where the 
     coordinate changes, and \textit{alien derivatives constraint}, respectively, which is a consequence of the fact that derivatives with respect to other coordinates than the one along the 
     coordinate line should be treated as independent variables. 
     Since the multiform principle implies that the criticality must hold for all time-curves, the system \eqref{gEL}  must hold simultaneously. 
    
    The set of equations \eqref{eq:gEL}, for a given set of Lagrangian components $L_i$, can be considered as yielding 
    a system of ordinary differential equations for the variables $\mathbf{X}(t_1,t_2,\cdots)$ 
    as functions of the multiple time variables $t_i$, which is compatible (hence, integrable) 
    as a system of commuting flows provided the system is compatible. 
    However this will impose conditions on those components $L_i$ which 
    can no longer (unlike in conventional Lagrangian theory) be freely 
    chosen. Thus, we will take here a different point of view from the 
    conventional one, namely to 
    consider the entire system \eqref{gEL} as a system of equations for the Lagrangian 
    components $L_i$ themselves, from which the latter needs to be solved. In general this amounts to solving a complicated system of 
    PDEs for the $L_i$, which is almost impossible without making 
    some simplifying assumptions on the form of these Lagrangian components. However, if one only specifies how these components depend on the derivatives 
    $\partial\mathbf{X}/\partial t_i$, i.e. assuming the kinetic form of the Lagrangians while leaving the  
    potentials (i.e. the coefficients depending only on $\mathbf{X}$ itself) free, then we will show that in the case of Calogero-Moser type systems (with long-range potentials only depending on the differences $X_i-X_j$) the generalised system of Euler-Lagrange equations \eqref{gEL} together with the closure predict the form of the potentials almost uniquely, 
    leading to the well-known integrable cases. More generally, within the choice of kinetic form, the 
    multiform problem leads to a classification problem that can be solved in 
    explicit form. 

    We finish this general discussion of the 1-form principle with the  that observation that the set of gEL equations, can be compactly 
    expressed in the language of a variational bicomplex (cf. e.g. \cite{Dickey}), namely as the simple condition 
    $\delta\mathrm{d}\mathscr{L} = 0$, where  $\mathrm{d}$ is the exterior derivative and $\delta$ is the variational derivative, cf. \cite{Suris_2016}, which follows from the \textit{pluri-Lagrangian principle} that 
    $S$ must be simultaneously critical for all choices of surfaces, 
    in particular closed surfaces to which one can apply a generalised 
    Stokes' theorem.

     \subsection{Deriving integrable CM type systems from the 1-form principle}\label{ssection2.2}
     
     In the previous section, important ingredients, including the gEL equation and the closure relation, have been introduced. In this section, we will treat Lagrangians as a set of solutions satisfying those compatible equations. We shall explore a system of $N$-body in one dimension with a long-range interaction known as the Caloger-Moser type systems and, for the sake of simplicity, only the first two flows will be studied.
     
     First, we take an \textit{ansatz} form for the first two Lagrangians as follows:
     \\
     \textbf{Lagrangian Ansatz: The CM system}

     \begin{equation}
         L_2 = \sum_{i=1}^N\frac{1}{2}\left(\frac{\partial X_i}{\partial t_2}\right)^2+\sum_{i\neq j}^NV(X_i-X_j)\;,\label{cm_L2}
     \end{equation}
      and,
     \begin{equation}
         L_3=\sum_{i=1}^N\left[\frac{\partial X_i}{\partial t_3}\frac{\partial X_i}{\partial t_2}+\alpha\left(\frac{\partial X_i}{\partial t_2}\right)^3\right]+\sum_{i\neq j}^N\frac{\partial X_i}{\partial t_2}\;W(X_i-X_j)\;.
     \end{equation}
      Note that the labelling of the Lagrangian components, and the corresponding time-variables, for the CM system is motivated by the fact that in the CM hierarchy the 
      first nontrivial component is of second degree, see \cite{RinCM}. Without loss of generality, we can assume the potentials $V$ and $W$ to be even functions of their arguments. 
     \\
     \\
     \textbf{Lagrangian Ansatz: The GF system}
     \begin{equation}
        L_1 = \sum_{i=1}^N\frac{\partial X_i}{\partial t_1}\ln{\abs{\frac{\partial X_i}{\partial t_1}}}+\sum_{i\neq j}^N\frac{\partial X_i}{\partial t_1}\;V(X_i-X_j)\;,\label{gf_L1}
    \end{equation}
    and
    \begin{equation}
        L_2 = \sum_{i=1}^N\left(\frac{\partial X_i}{\partial t_2}\ln{\abs{\frac{\partial X_i}{\partial t_1}}}+\beta\frac{\partial X_i}{\partial t_2}\right)+\sum_{i\neq j}^N\left(\frac{\partial X_i}{\partial t_2}\;W(X_i-X_j)+\frac{\partial X_i}{\partial t_1}\frac{\partial X_j}{\partial t_1}\;U(X_i-X_j)\right)\;.\label{gf_L2}
    \end{equation}
    We note that the $L_2$ is a bit different from the one given in \cite{Jairuk1} in order to get a right constraint. We shall also assume the potentials $V$, $W$ and $U$ to be even functions of their arguments.
    \\
    \\
    \textbf{Lagrangian Ansatz: The RS system}
    \begin{equation}
        L_1 = \sum_{i=1}^N\frac{\partial X_i}{\partial t_1}\ln{\abs{\frac{\partial X_i}{\partial t_1}}}+\sum_{i\neq j}^N\frac{\partial X_i}{\partial t_1}\;V_\lambda(X_i-X_j)\;,\label{rs_L1}
    \end{equation}
    and,
    \begin{equation}
        L_2 = \sum_{i=1}^N\left(\frac{\partial X_i}{\partial t_2}\ln{\abs{\frac{\partial X_i}{\partial t_1}}}+\frac{1}{\lambda}\left(\frac{\partial X_i}{\partial t_1}\right)^2+\beta\frac{\partial X_i}{\partial t_2}\right)+\sum_{i\neq j}^N\left(\frac{\partial X_i}{\partial t_2}\;W_\lambda(X_i-X_j)+\frac{\partial X_i}{\partial t_1}\frac{\partial X_j}{\partial t_1}\;U_\lambda(X_i-X_j)\right)\;.
    \end{equation}
    Here, a parameter $\lambda$ will play an important role at certain limits in obtaining the CM and GF systems. We again shall also assume the potentials $V_\lambda$, $W_\lambda$ and $U_\lambda$ to be even functions of their arguments.
    \\
    \\
    \\
   Second, we compute the equations of motion and constraint for each system as follows:
    \\
    \textbf{The CM system}\\
    The first equation for $t_2$ is 
    \begin{equation}
         \frac{\partial^2X_i}{\partial t_2^2} = 2\sum_{\substack{j=1\\j\neq i}}^NV'(X_i-X_j)\;,\label{Lcm_EoM2}
     \end{equation}
     and the equation of motion for $t_3$-component is
     \begin{equation}
         \frac{\partial^2X_i}{\partial t_2\partial t_3} = \sum_{\substack{j=1\\j\neq i}}^N\left[W'(X_i-X_j)\frac{\partial X_i}{\partial t_2}-W'(X_j-X_i)\frac{\partial X_j}{\partial t_2}\right]\;,\label{Lcm_EoM3}
     \end{equation}
     where $V'$ and $W'$ denote the derivative of the functions with respect to their arguments. The constraint is given by
     \begin{equation}
         \frac{\partial L_3}{\partial\left(\frac{\partial X_i}{\partial t_2}\right)} = 3\alpha\left(\frac{\partial X_i}{\partial t_2}\right)^2+\frac{\partial X_i}{\partial t_3}+\sum_{\substack{j=1\\j\neq i}}^NW(X_i-X_j)=0\;.\label{Lcm_cn3}
     \end{equation}
    \\
    \textbf{The GF system}\\
    The first equation reads
    \begin{equation}
        \frac{\partial^2X_i}{\partial t_1^2} = \frac{\partial X_i}{\partial t_1}\sum_{\substack{j=1\\j\neq i}}^N\frac{\partial X_j}{\partial t_1}\left[V'(X_i-X_j)-V'(X_j-X_i)\right]\;,\label{Lgf_EoM1}
    \end{equation}
    and the second one is
    \begin{equation}
        \frac{\partial^2X_i}{\partial t_1\partial t_2} = \frac{\partial X_i}{\partial t_1}\sum_{\substack{j=1\\j\neq i}}^N\left(\frac{\partial X_j}{\partial t_2}\left[W'(X_i-X_j)-W'(X_j-X_i)\right]+\frac{\partial X_j}{\partial t_1}\left[U'(X_i-X_j)-U'(X_j-X_i)\right]\right)\;,\label{Lgf_EoM2}
    \end{equation} 
    where $V'$, $W'$, and $U'$ denote the derivative of the functions with respect to their arguments.
     The constraint is given by
    \begin{equation}
        \frac{\partial L_2}{\partial\left(\frac{\partial X_i}{\partial t_1}\right)} = \frac{\partial X_i}{\partial t_2}\bigg/\frac{\partial X_i}{\partial t_1}+\sum_{\substack{j=1\\j\neq i}}^N\frac{\partial X_j}{\partial t_1}\left[U(X_i-X_j)+U(X_j-X_i)\right]=0\;.\label{Lgf_cn}
    \end{equation}
    Additionally, the corner equation provides
    \begin{equation}
        \frac{\partial L_1}{\partial(\partial X_i/\partial t_1)} = \frac{\partial L_2}{\partial(\partial X_i/\partial t_2)}\;\;\;\implies\;\;\;V(X_i-X_j) = W(X_i-X_j)+\beta-1\;,\label{Lgf_cor}
    \end{equation}
    for all $i\neq j$.
    \\
    \\
    \textbf{The RS system}\\
    The first equation is 
    \begin{equation}
        \frac{\partial^2X_i}{\partial t_1^2} = \frac{\partial X_i}{\partial t_1}\sum_{\substack{j=1\\j\neq i}}^N\frac{\partial X_j}{\partial t_1}\left[V'_\lambda(X_i-X_j)-V'_\lambda(X_j-X_i)\right]\;,\label{Lrs_EoM1}
    \end{equation}
    and the second equation is
    \begin{equation}
        \frac{\partial^2X_i}{\partial t_1\partial t_2} = \frac{\partial X_i}{\partial t_1}\sum_{\substack{j=1\\j\neq i}}^N\left(\frac{\partial X_j}{\partial t_2}\left[W'_\lambda(X_i-X_j)-W'_\lambda(X_j-X_i)\right]+\frac{\partial X_i}{\partial t_1}\frac{\partial X_j}{\partial t_1}\left[U'_\lambda(X_i-X_j)-U'_\lambda(X_j-X_i)\right]\right)\;,\label{Lrs_EoM2}
    \end{equation}
     where $V'_\lambda$, $W'_\lambda$ and $U'_\lambda$ denote the derivative of the functions with respect to their arguments. The constraint is
    \begin{equation}
        \frac{\partial L_2}{\partial\left(\frac{\partial X_i}{\partial t_1}\right)} = \frac{2}{\lambda}\frac{\partial X_i}{\partial t_1}+\frac{\partial X_i}{\partial t_2}\bigg/\frac{\partial X_i}{\partial t_1}+\sum_{\substack{j=1\\j\neq i}}^N\frac{\partial X_j}{\partial t_1}\left[U_\lambda(X_i-X_j)+U_\lambda(X_j-X_i)\right]=0\;.\label{Lrs_cn}
    \end{equation}
    And, what we obtain from the corner equation is
    \begin{equation}
        V_\lambda(X_i-X_j) = W_\lambda(X_i-X_j)+\beta-1\;,\label{Lrs_cor}
    \end{equation}
    for all $i\neq j$.
    \\
    \\
    
    Now that we have all ingredients, we are ready to solve for potential functions as follows:
    \\
    \textbf{The CM system}
    \\
    One takes its derivative with respect to $t_2$ on \eqref{Lcm_cn3} and eliminates the second-order derivatives by using \eqref{Lcm_EoM2} and \eqref{Lcm_EoM3}, resulting in
     \begin{equation}
         W(x) = -6\alpha V(x)+\text{constant}\;.\label{Lcm_VW}
     \end{equation}
     We recall the closure relation \eqref{closure} for first two flows:
     $\frac{\partial L_2}{\partial t_3}=\frac{\partial L_3}{\partial t_2},$
     and consider this as a \textit{on-shell} condition, i.e., that it holds only on the solution of the equation of motion. Then employing \eqref{Lcm_EoM2}, \eqref{Lcm_EoM3}, \eqref{Lcm_cn3}, the closure relation gives
     \begin{equation}
         \sum_{i\neq j\neq k}^NW'(X_i-X_j)W(X_j-X_k) = 0\;.\label{Lcm_cyclicell}
     \end{equation}
     In the particular case of $N=3$, we get a functional relation in terms of free particle $x=X_1-X_2$, $y=X_2-X_3$:
     \begin{equation}
            W'(x)\left[W(y)-W(z)\right]-W'(y)\left[W(x)-W(z)\right]=W'(z)\left[W(x)-W(y)\right]\;,\label{Lm_addell}
    \end{equation}
    where $z=x+y$. We recognise that the relation \eqref{Lm_addell} is a well-known addition formula for the Weierstrass $\wp$-function.
    The general solution of \eqref{Lm_addell} is given by
    \begin{equation}
         W(x) = C_1\wp(x)+C_2\;,\label{Lcm_sol}
    \end{equation}
    where $C_1$ and $C_2$ are constant. With \eqref{Lcm_sol}, the potential is determined up to the unimportant additive constant $C_2$, and hence this fixes the potentials in the Lagrangian system uniquely. Then, this corresponds to the well-known elliptic Calogero-Moser case which is a famous integrable equation.
    \\
    \\
    \textbf{The GF system}
    \\
We take the derivative with respect to $t_2$ on \eqref{Lgf_cn} and eliminate the second-order derivatives by using \eqref{Lrs_EoM1} and \eqref{Lrs_EoM2}, resulting in
     \begin{equation}
         \sum_{\substack{j=1\\j\neq i}}^N\left[A(X_i-X_j)\left(\frac{\partial X_i}{\partial t_1}\right)^2\frac{\partial X_j}{\partial t_1}+B(X_i-X_j)\frac{\partial X_i}{\partial t_1}\left(\frac{\partial X_j}{\partial t_1}\right)^2\right]+\sum_{\substack{j\neq k\\j,k\neq i}}^NC(X_i-X_j,X_j-X_k)\frac{\partial X_i}{\partial t_1}\frac{\partial X_j}{\partial t_1}\frac{\partial X_k}{\partial t_1} = 0\;,\label{Lgf_compat}
     \end{equation}
     where, in the particular case of $N=3$,
    \begin{subequations}
        \begin{align}
             A(x) =& \left[U(x)+U(-x)\right]\left[W'(-x)-W'(x)+V'(-x)-V'(x)\right]+2\left[U'(x)-U'(-x)\right]\;,\\
             B(x) =& \left[U(x)+U(-x)\right]\left[V'(x)-V'(-x)\right]-\left[U'(x)-U'(-x)\right]\;,\\
             C(x,y) =& \Big(U(x)+U(-x)\Big)\Big(V'(y)-V'(-y)+V'(z)-V'(-z)\Big) \nonumber
         \\&+ \Big(U(z)+U(-z)\Big)\Big(V'(x)-V'(-x)-V'(y)+V'(-y)\Big) \nonumber
         \\&-\Big(U(y)+U(-y)\Big)\Big(W'(x)-W'(-x)+W'(z)-W'(-z)\Big)\;.
        \end{align}
    \end{subequations}
     Using \eqref{Lgf_cor}, one chooses $A=0$ and $B=0$ leading to
     the relation
     \begin{equation}
         \Phi(x) := V(x)+V(-x) = \ln\left(U(x)+U(-x)\right)+\text{constant}\;. \label{LgfVUrela}
     \end{equation}
     The condition $C=0$ leads to the following functional relation for the potentials
     providing
     \begin{equation}
         e^{\Phi(x)}\left(\Phi'(y)+\Phi'(z)\right)-e^{\Phi(y)}\left(\Phi'(x)+\Phi'(y)\right)=e^{\Phi(z)}\left(\Phi'(y)-\Phi'(x)\right)\;.
     \end{equation}
     A solution of this functional equation, as a consequence of the addition formula \eqref{Lm_addell}, is given by 
     \begin{equation}
         \Phi(x) = -\ln|C_1\wp(x)+C_2|\;,\label{Lgf_sol}
     \end{equation}
     where $C_1$ and $C_2$ are constant.
      \\
      \\
      With some effort, it can be proven that it is indeed a general solution, which upon inserting $\Phi(x)$ back to find the potentials, leading to an elliptic case of the GF flows. Thus we see in this case that only gEL equation leads to the selection of the elliptic integrability. We still need to verify the closure relation, as we shall see, is automatically satisfied under the above choice of potentials. Recalling the closure relation \eqref{closure} for the first two flows,
     $\frac{\partial L_1}{\partial t_2}=\frac{\partial L_2}{\partial t_1}\;$, we consider this as condition \textit{on-shell}, i.e., that it holds only on the solution of the equation of motion. Thus, using \eqref{Lgf_EoM1}-\eqref{Lgf_cor}, and inserting into the closure relation, we find again the relation \eqref{Lgf_compat}.
    \\
    \\
    \textbf{The RS system}
    \\
    We again take the derivative with respect to $t_2$ and eliminate the second-order derivatives by using \eqref{Lrs_EoM1} and \eqref{Lrs_EoM2}, resulting in
     \begin{eqnarray}
         &\sum_{\substack{j=1\\j\neq i}}^N\left[A_\lambda(X_i-X_j)\left(\frac{\partial X_i}{\partial t_1}\right)^2\frac{\partial X_j}{\partial t_1}+ B_\lambda(X_i-X_j)\frac{\partial X_i}{\partial t_1}\left(\frac{\partial X_j}{\partial t_1}\right)^2\right]\nonumber\\
         &\;\;\;\;\;\;\;\;\;\;\;\;\;\;\;\;\;\;\;\;\;\;\;\;\;\;\;\;\;\;\;\;\;\;\;\;\;\;\;\;\;\;\;\;\;\;\;\;\;\;\;\;\;\;\;\;\;\;\;+\sum_{\substack{j\neq k\\j,k\neq i}}^NC_\lambda(X_i-X_j,X_j-X_k)\frac{\partial X_i}{\partial t_1}\frac{\partial X_j}{\partial t_1}\frac{\partial X_k}{\partial t_1} = 0\;,\label{Lrs_compat}
      \end{eqnarray}
     where (again in the case of $N=3$)
     \begin{subequations}
        \begin{align}
             A_\lambda(x) =& -\left[U_\lambda(x)+U_\lambda(-x)\right]\left[W'_\lambda(x)-W'_\lambda(-x)+V'_\lambda(x)-V'_\lambda(-x)\right]+2\left[U'_\lambda(x)-U'_\lambda(-x)\right]\nonumber
             \\&+\frac{4}{\lambda}\left[V'_\lambda(x)-V'_\lambda(-x)\right] \;,\\
             B_\lambda(x) =& -\left[U'_\lambda(x)-U'_\lambda(-x)\right]-\frac{2}{\lambda}\left[W'_\lambda(x)-W'_\lambda(-x)\right]+\left[U_\lambda(x)+U_\lambda(-x)\right]\left[V'_\lambda(x)-V'_\lambda(-x)\right]\;,\\     
             C_\lambda(x,y) =& \Big(U_\lambda(x)+U_\lambda(-x)\Big)\Big(V'_\lambda(y)-V'_\lambda(-y)+V'_\lambda(z)-V'_\lambda(-z)\Big)\nonumber
         \\&+ \Big(U_\lambda(z)+U_\lambda(-z)\Big)\Big(V'_\lambda(x)-V'_\lambda(-x)-V'_\lambda(y)+V'_\lambda(-y)\Big) \nonumber
         \\&-\Big(U_\lambda(y)+U_\lambda(-y)\Big)\Big(W'_\lambda(x)-W'_\lambda(-x)+W'_\lambda(z)-W'_\lambda(-z)\Big)\;.
        \end{align}
    \end{subequations}
     By inserting \eqref{Lrs_cor} into the conditions $A_\lambda=0$ and $B_\lambda=0$, we obtain the relation:
     \begin{equation}
         \Phi_\lambda(x) := V_\lambda(x)+V_\lambda(-x) = \ln\left(U_\lambda(x)+U_\lambda(-x)-\frac{2}{\lambda}\right)+\text{constant}\;.\label{LrsVUrela}
     \end{equation}
     Here, we take an \textit{ansazt} for the parametric potential as
     \begin{equation}
            V'_\lambda(x) = \frac{V'(x)}{V(x)\left(1-\frac{V(x)}{V(\lambda)}\right)}\;,\label{Lrs_v_lam}
     \end{equation}
     with the GF limit:  $\lim_{\lambda\to\infty}1/V(\lambda) = 0$. The condition $C_\lambda=0$ leads to the following functional relation for the potentials resulting in
     \begin{equation}
         e^{\Phi_\lambda(x)}\left(\Phi'_\lambda(y)+\Phi'_\lambda(z)\right)-e^{\Phi_\lambda(y)}\left(\Phi'_\lambda(x)+\Phi'_\lambda(y)\right)=e^{\Phi_\lambda(z)}\left(\Phi'_\lambda(y)-\Phi'_\lambda(x)\right)\;.\label{Lrs_func}
     \end{equation}
     According to the addition formula \eqref{Lm_addell}, a solution of this functional equation, corresponding to the formula \eqref{Lrs_v_lam}, is given by 
     \begin{equation}
         \Phi_\lambda(x) = -\ln\abs{C_1(\wp(\lambda)-\wp(x))+C_2}\;,\label{Lrs_sol}
     \end{equation}
     where $C_1$ and $C_2$ are constant. The potential \eqref{Lrs_sol} is determined up to the unimportant additive constant $C_2$, and hence this fixes the potentials in the Lagrange system uniquely. It is not difficult to see that \eqref{Lrs_sol} satisfies the Goldfish limit $\lambda \rightarrow \infty$. With \eqref{Lrs_sol}, thus, this corresponds to the well-known elliptic RS case. Therefore, one can insert $\Phi_\lambda(x)$ back to the potentials for the proof that it is indeed a general solution leading to an elliptic case of the RS flows. Again, only gEL equation leads to the selection of the elliptic integrability. Moreover, the closure relation $\frac{\partial L_1}{\partial t_2}=\frac{\partial L_2}{\partial t_1}\;,$ has to be still verified \textit{on-shell}, i.e., that it holds only on the solution of the equation of motion. Hence, using \eqref{Lrs_EoM1}-\eqref{Lrs_cor}, and inserting into the closure relation, we find again the relation \eqref{Lrs_compat}
     \\
     \\
     To summarise, the ans\"atze for the kinetic terms in the Lagrangian 1-forms for the CM, GF, and RS systems, comprising 
     only two time-flows in the corresponding hierarchies, are sufficient to uniquely specify the most general integrable potentials (the elliptic potential) 
     from the system of Euler-Lagrange equations, i.e., gEL equation and closure relation.

\section{Hamiltonian 1-form structure}\label{section3} 


    In this section, we will examine the alternative perspective namely the Hamiltonian 1-forms. The Hamiltonian form of the least action principle was formulated in 
    \cite{Caudrelier_2020}. 

    \subsection{A brief review of Hamiltonian integrability}\label{ssection3.1}
    
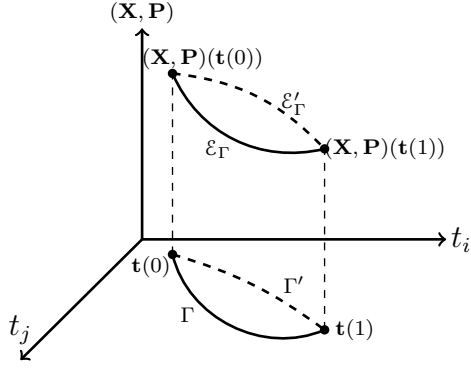
\begin{figure}[h]
    	    \centering
        	    \begin{tikzpicture}[scale = 0.8]
            	    \path[line width = 1pt, draw = black,->]
 	                  (0,0) to (5,0);
 	                  
 	                  \path[line width = 1pt, draw = black,->]
 	                  (0,0) to (0,3.5);

                        \path[line width = 1pt, draw = black,->]
 	                  (0,0) to (-2,-2);
 	                  
 	                \path[line width = 1pt, draw = black]
 	                  (0.5,-0.25) to [bend right = 50] (3,-1.5);

                        \path[dashed, line width = 1pt, draw = black]
 	                  (0.5,-0.25) to [bend left = 10] (3,-1.5);
 	                 
 	                \path[line width = 1pt, draw = black]
 	                  (0.5,2.75) to [bend right = 40] (3,1.5);

                        \path[dashed, line width = 1pt, draw = black]
 	                  (0.5,2.75) to [bend left = 20] (3,1.5);
                    
                        \path[dashed, line width = 0.5pt, draw = black]
 	                  (0.5,-0.25) to (0.5,2.75);
                        \path[dashed, line width = 0.5pt, draw = black]
 	                  (3,-1.5) to (3,1.5);
                    
 	                 \node at (5.25,0) {$t_i$};
 	                 \node at (-2,-1.5) {$t_j$};
 	                 \node at (0,3.75) {\scriptsize{$(\mathbf{X},\mathbf{P})$}};

                        \node[circle, fill, inner sep=1.25 pt] at (0.5,-0.25) {};
                        \node[circle, fill, inner sep=1.25 pt] at (3,-1.5) {};
 	                \node at (0.15,-0.45) {\scriptsize{$\mathbf{t}(0)$}};
 	                \node at (3.5,-1.5) {\scriptsize{$\mathbf{t}(1)$}};

                        \node[circle, fill, inner sep=1.25 pt] at (0.5,2.75) {};
                        \node[circle, fill, inner sep=1.25 pt] at (3,1.5) {};
 	                \node at (1,3) {\scriptsize{$(\mathbf{X},\mathbf{P})(\mathbf{t}(0))$}};
 	                \node at (4,1.5) {\scriptsize{$(\mathbf{X},\mathbf{P})(\mathbf{t}(1))$}};
                   
 	                 \node at (2.5,-0.75) {\scriptsize{$\Gamma'$}};
 	                 \node at (0.75,-1.25) {\scriptsize{$\Gamma$}};
                        \node at (1.25,1.5) {\scriptsize{$\mathcal{E}_\Gamma$}};
 	                 \node at (2.5,2.25) {\scriptsize{$\mathcal{E}'_\Gamma$}};
 	                 
 	            \end{tikzpicture}
              \captionof{figure}{The curves $\Gamma$ and $\mathcal{E}_\Gamma$ in the $(\mathbf{X},\mathbf{P})-\mathbf{t}$ configuration.}\label{fig2}
        \end{figure}

    Consider again a system of $N$-particles described by a set of positions $\mathbf{X} = \{X_1,...,X_N\}$ on the parametrised curve $\Gamma$, given in Fig.\ref{fig2}, of time $\mathbf{t}(s)=\{t_1(s),...,t_N(s)\}$,  where $s$ is a parametrised variable: $s\in\left[0,1\right]$. We first start to establish the Legendre transformation by searching the conserved quantity, which is analogous to the standard Hamiltonian being the conserved quantity under time translation, corresponding to $s$(parametrised variable)-flow \cite{Puttarprom2019IntegrableHH, Suris1}. Here, the momentum variable is defined as well as the action functional. Then, the Hamiltonian commuting flows can be obtained from the variational principle with respect to independent variable space. Moreover, the variational principle, applied to the action in phase space, provides a pair of Hamilton equations. 
    \\
    \\
    We first introduce the relation
    \begin{equation}
        \frac{d}{ds}\left(\sum_{i=1}^NL_i\frac{dt_i}{ds}\right) = \sum_{i\neq j}^N\left(\frac{\partial L_i}{\partial \mathbf{X}}\frac{d\mathbf{X}}{ds}+\frac{\partial L_i}{\partial \left(\partial\mathbf{X}/\partial t_j\right)}\frac{d\left(\partial\mathbf{X}/\partial t_j\right)}{ds}\right)\frac{dt_i}{ds}+\sum_{i=1}^NL_i\frac{d^2t_i}{ds^2}\;.\label{H_dL/ds}
    \end{equation}
     Imposing $d^2t_i/ds^2=0$ and using the EL equation \eqref{gEL} as well as the constraints \eqref{corner}, equation \eqref{H_dL/ds} is given by
     \begin{equation}
         0=\frac{d}{ds}\left[\frac{1}{N}\frac{d\mathbf{X}}{ds}\left(\sum_{i=1}^N\frac{\partial L_i}{\partial \left(\partial\mathbf{X}/\partial t_i\right)}+\sum_{i\neq j}^N\frac{\partial L_i}{\partial \left(\partial\mathbf{X}/\partial t_j\right)}\frac{dt_i/ds}{dt_j/ds}\right)-\sum_{i=1}^NL_i\frac{dt_i}{ds}\right]\;.
     \end{equation}
     The momentum variable is defined by
     \begin{equation}
         \mathbf{P} := \frac{1}{N}\left(\sum_{i=1}^N\frac{\partial L_i}{\partial \left(\partial\mathbf{X}/\partial t_i\right)}+\sum_{i\neq j}^N\frac{\partial L_i}{\partial \left(\partial\mathbf{X}/\partial t_j\right)}\frac{dt_i/ds}{dt_j/ds}\right)\;.\label{PL}
     \end{equation}
     Therefore, in terms of Hamiltonian components, the action functional, evaluated on the curve $\Gamma$, \eqref{action} becomes
     \begin{equation}
         S[\mathbf{X}(\mathbf{t}),\mathbf{P}(\mathbf{t}),\Gamma] = \int_\Gamma\;\sum_{i=1}^N\left(\mathbf{P}\cdot\frac{\partial \mathbf{X}}{\partial t_i}-H_i\right)dt_i = \int_0^1\;\sum_{i=1}^N\left[\left(\mathbf{P}\cdot\frac{\partial \mathbf{X}}{\partial t_i}-H_i\right)\frac{dt_i}{ds}\right]ds\;,\label{Haction}
     \end{equation}
     where the term inside the bracket is the Legendre transformation for obtaining the Hamiltonian hierarchy. 
        \\
        \\
         Next, we now perform the variation on the field variables $(\mathbf{X},\mathbf{P})$. We now fix the curve $\Gamma$ on the independent variable space and examine the corresponding evaluated curve $\mathcal{E}_\Gamma$ in the extended space of independent variables $\mathbf{t}$ and field variables $(\mathbf{X},\mathbf{P})$ given in Fig.\ref{fig2}. The variation of action \eqref{Haction}, subjects to the variation of field variables $(\mathbf{X},\mathbf{P})\mapsto(\mathbf{X}+\delta\mathbf{X},\mathbf{P}+\delta\mathbf{P})$, i.e., $\mathcal{E}_\Gamma\mapsto\mathcal{E}'_\Gamma$, given in Fig.\ref{fig2}, with the boundary conditions $\delta\mathbf{X}(\mathbf{t}(0))=\delta\mathbf{X}(\mathbf{t}(1))=0$, is given by 
     \begin{eqnarray}
         \delta S = \int_0^1ds\sum_{i=1}^N\left[\delta\mathbf{P}\left(\frac{d\mathbf{X}}{ds}-\frac{\partial H_i}{\partial\mathbf{P}}\frac{dt_i}{ds}\right)-\delta\mathbf{X}\left(\frac{d\mathbf{P}}{ds}+\frac{\partial H_i}{\partial\mathbf{X}}\frac{dt_i}{ds}\right)\right]\;.
     \end{eqnarray}
     Imposing a critical condition $\delta S = 0$, the coefficients of $\delta\mathbf{X}$ and $\delta\mathbf{P}$ vanishes, one obtains
         \begin{equation}
             \frac{d\mathbf{X}}{ds} = \sum_{i=1}^N\frac{\partial H_i}{\partial\mathbf{P}}\frac{dt_i}{ds}\;,\;\;\;\text{and}\;\;\;\frac{d\mathbf{P}}{ds} = -\sum_{i=1}^N\frac{\partial H_i}{\partial\mathbf{X}}\frac{dt_i}{ds}\;,\label{gHeq}
         \end{equation} 
     which is called a set of \textit{generalised Hamilton equations}. Next, one considers the derivative of the component inside the bracket of \eqref{Haction}, resulting in
     \begin{equation}
         d\mathscr{L} = \sum_{i<j}^N\left[\left(\frac{\partial\mathbf{P}}{\partial t_j}\cdot\frac{\partial \mathbf{X}}{\partial t_i}-\frac{\partial\mathbf{P}}{\partial t_i}\cdot\frac{\partial \mathbf{X}}{\partial t_j}\right)-\left(\frac{\partial H_i}{\partial t_j}-\frac{\partial H_j}{\partial t_i}\right)\right]dt_j\wedge dt_i\;.\label{deLHrela}
     \end{equation}
     Here, we introduce the symplectic matrix and symplectic coordinate such that
     \begin{equation}
        \Omega = 
         \begin{pmatrix}
            \mathbf{0}_{N\times N} & \mathbb{I}_{N\times N}\\
            -\mathbb{I}_{N\times N} & \mathbf{0}_{N\times N}
            \end{pmatrix}_{2N\times 2N}\;\;\text{and}\;\;
            \Lambda = 
            \begin{pmatrix}
                \mathbf{X}_{N\times 1 }\\
                \mathbf{P}_{N\times 1}
            \end{pmatrix}_{2N\times 1}\;.\label{omegalambda}
     \end{equation}
     where $\mathbf{0}_{N\times N}$ is an zero matrix and $\mathbb{I}_{N\times N}$ is an identity matrix. Now, the equation \eqref{deLHrela} can be rewritten as
     \begin{align}
         d\mathscr{L} &= \sum_{i<j}^N\left[\frac{\partial\Lambda^T}{\partial t_i}\Omega\frac{\partial\Lambda}{\partial t_j}-\left(\frac{\partial H_i}{\partial\Lambda}\right)^T\frac{\partial\Lambda}{\partial t_j}+\frac{\partial \Lambda^T}{\partial t_i}\frac{\partial H_j}{\partial \Lambda}\right]\; dt_j\wedge dt_i\;,\nonumber
         \\&=\sum_{i<j}^N\left\{-\left[\Omega\frac{\partial \Lambda}{\partial t_i}+\frac{\partial H_i}{\partial\Lambda}\right]^T\Omega^{-1}\left[\Omega\frac{\partial \Lambda}{\partial t_j}+\frac{\partial H_j}{\partial \Lambda}-\frac{\partial H_j}{\partial \Lambda}\right]+\frac{\partial \Lambda^T}{\partial t_i}\frac{\partial H_j}{\partial \Lambda}\right\}\; dt_j\wedge dt_i\;,\nonumber
         \\&=\sum_{i<j}^N\left(-\gamma_i^T\Omega^{-1}\gamma_j+\{H_j,H_i\}\right)\; dt_j\wedge dt_i\;,\label{LHrela}
     \end{align}
     which is identically provided in \cite{caudrelier2024lagrangian}, where 
     \begin{subequations}
         \begin{equation}
             \gamma_i = \Omega\frac{\partial \Lambda}{\partial t_i}+\frac{\partial H_i}{\partial \Lambda}\;,\label{LHA}
         \end{equation}
         \begin{equation}
             \{H_i,H_j\} = \left(\frac{\partial H_j}{\partial \Lambda}\right)^T\Omega^{-1}\frac{\partial H_i}{\partial \Lambda}\;.\label{LHB}
         \end{equation}
     \end{subequations}
     Equation \eqref{LHrela} is  the ``double zero" formula for the exterior derivative of the Lagrangian 1-form, that as a genereal 
     phenomenon was highlighted in \cite{Sleigh_2019_2} in the context of 
     variational symmetries and Lagrangian 2-forms. In the context of Hamiltonian dynamics, the integrability can be identified by \eqref{LHrela} with \textit{on-shell} condition consequently vanishing of \eqref{LHA}, $\gamma_i = 0$, such that
     \begin{eqnarray}
         \left\{H_i,H_j\right\} = 0\;,\;\;\text{for}\;\;i,j=1,2,...,N\;.\label{Hcommute}
     \end{eqnarray}
     To summarise, performing the least action principle applied to the curve $\Gamma$ in the space of independent variables, the computation leads to the Hamiltonian commuting flows \eqref{Hcommute}. While fixing the curve $\Gamma$ and varying the curve $\mathcal{E}_\Gamma$, the computation leads to the generalised Hamilton equations corresponding to Hamiltonian 1-form. 
\\
\\
       We would like to mention that recently an important connection was established between the Lagrangian multiforms and the $r$-matrix, namely the classical Yang-Baxter equation (CYBE), employing some particular models e.g. the open Toda system and the rational Gaudin model \cite{Caudrelier_2024, caudrelier2024lagrangian} 
     \subsection{Re-deriving the CM cases from the Hamilton 1-form principle}\label{ssection3.2}
     In the previous section, important ingredients, including the generalised Hamilton equations and the Hamiltonian commuting flows, have been introduced for testing the integrability for Hamiltonian 1-forms. Here, in this section, we will treat Hamiltonian as a set of solutions satisfying these compatible equations. Then, it means that a system, possessing a set of Hamiltonians obeying these compatible equations, is integrable. We shall again explore a system of $N$-body in one dimension with a long-range interaction known as the Calogero-Moser type systems and, for the sake of simplicity, only the first two flows will be studied.
     \\
     \\
     $\bullet$ We take an \textit{ansatz} form for the first two Hamitonians, given by Legendre transformation expressed in components of \eqref{Haction}, as follows:
     \\
     \textbf{The CM system}
        \begin{equation}
             H_2 = \sum_{i=1}^N \frac{P_i^2}{2}+\sum_{i\neq j}^N V(X_i-X_j)\;,
        \end{equation}
        and,
        \begin{equation}
            H_3 =  \sum_{i=1}^N \frac{P_i^3}{4}+\sum_{i\neq j}^N P_iW(X_i-X_j)\;,
        \end{equation}
        where $V$ and $W$ are even functions of their arguments.
     \\
     \\
     \textbf{The GF system}
    \begin{equation}
         H_1 = \sum_{i=1}^Ne^{P_i+\sum_{\substack{j=1\\j\neq i}}^NV(X_i-X_j)}\;,\label{gf_H1}
    \end{equation}
    and,
    \begin{equation}
         H_2 = \sum_{i=1}^Ne^{P_i+\sum_{\substack{j=1\\j\neq i}}^NW(X_i-X_j)}\sum_{\substack{l=1\\l\neq i}}^Ne^{P_l+\sum_{\substack{j=1\\j\neq l}}^NW(X_l-X_j)}U(X_i-X_l)\;.\label{gf_H2}
    \end{equation}
    We note that the $H_2$ is modified a bit from the one given in \cite{Suris1} to get the right constraint. And, $V$, $W$ and $U$ are even function of their arguments.
    \\
    \\
    \textbf{The RS system}
    \begin{equation}
        H_1 = \sum_{i=1}^Ne^{P_i+\sum_{\substack{j=1\\j\neq i}}^NV_\lambda(X_i-X_j)}\;,
    \end{equation}
    and,
    \begin{equation}
       H_2 = \sum_{i=1}^Ne^{P_i+\sum_{\substack{j=1\\j\neq i}}^NW_\lambda(X_i-X_j)}\bigg(\frac{1}{\lambda}e^{P_i+\sum_{\substack{j=1\\j\neq i}}^NW_\lambda(X_i-X_j)}+\sum_{\substack{l=1\\l\neq i}}e^{P_l+\sum_{\substack{j=1\\j\neq l}}^NW_\lambda(X_l-X_j)}U(X_i-X_l)\bigg)\;.
    \end{equation}
     Again, a parameter $\lambda$ will play an important role at certain limits in obtaining the CM and GF systems. We again shall also assume the potentials $V_\lambda$, $W_\lambda$, and $U_\lambda$ to be even functions of their arguments.
     \\
     \\
     $\bullet$ Next, we give the equations of motion obtaining from the generalised Hamilton equations for each system as follows:
     \\
    \textbf{The CM system}
    \\
     The generalised Hamilton equations provide
        \begin{align}
            \frac{\partial X_i}{\partial t_2} =& P_i\;,\label{cm_xdot2}
            \\\frac{\partial X_i}{\partial t_3} =& \frac{3}{4}P_i^2+\sum_{\substack{j=1\\j\neq i}}^NW(X_i-X_j)\;,\label{cm_xdot3}
            \\\frac{\partial P_i}{\partial t_2} =& -2\sum_{\substack{j=1\\j\neq i}}^NV'(X_i-X_j)\;,\label{cm_pdot2}
            \\\frac{\partial P_i}{\partial t_3} =&  -\sum_{\substack{j=1\\j\neq i}}^N\left(P_i+P_j\right)W'(X_i-X_j)\;,\label{cm_pdot3}
        \end{align}
        where $V'$ and $W'$ denote the derivative of the functions with respect to their arguments. Here we notice that, with \eqref{cm_xdot2}, the equation \eqref{cm_xdot3} and \eqref{cm_pdot3} are the constraints and the equation of motion for $t_3$-component given in \eqref{Lcm_cn3} and \eqref{Lcm_EoM3}, respectively.
        Using \eqref{cm_xdot2} and \eqref{cm_pdot2}, one obtains
        \begin{equation}
            \frac{\partial^2X_i}{\partial t_2^2} = -2\sum_{\substack{j=1\\j\neq i}}^NV'(X_i-X_j)\;,\label{Hcm_EoM2}
        \end{equation}
        which is the equation of motion for $t_2$-component. Alternatively, the equation of motion for $t_3$-component can be obtained by proceeding derivative $\partial/\partial t_2$ on \eqref{cm_pdot3} such that
        \begin{equation}
             \frac{\partial^2X_i}{\partial t_2 \partial t_3} = \frac{3}{2}P_i\frac{\partial P_i}{\partial t_2}+\sum_{\substack{j=1\\j\neq i}}^N\left(P_i-P_j\right)W'(X_i-X_j)\;.\label{Hcm_EoM3_2}
        \end{equation} 
        The equation \eqref{Hcm_EoM3_2} is equivalent to \eqref{cm_pdot3} and will be used later in solving the potential.
        \\
        \\
        \\
        \\
        \\
        \\
        \\
        \\
        \\
        \\
        \\
        \\
        \\
        \textbf{The GF system}
        \\
        The generalised Hamilton equations provide
        \begin{align}
            \frac{dX_i}{dt_1} =&\; e^{P_i+\sum_{\substack{j=1\\j\neq i}}^NV(X_i-X_j)}\;,\label{Hgf_Xdot_1}
            \\\frac{dX_i}{dt_2} =& \sum_{\substack{l=1\\l\neq i}}^N\frac{dX_i}{dt_1}\frac{dX_l}{dt_1}e^{\Psi^i+\Psi^l}\left[U(X_i-X_l)+U(X_l-X_i)\right]\;,\label{Hgf_Xdot_2}
            \\\frac{dP_i}{dt_1} =& -\frac{dX_i}{dt_1}\sum_{\substack{j=1\\j\neq i}}^NV'(X_i-X_j)+\sum_{\substack{j=1\\j\neq i}}^N\frac{dX_j}{dt_1}\;V'(X_j-X_i)\;,\label{Hgf_Pdot_1}
            \\\frac{dP_i}{dt_2} =& -\sum_{\substack{l=1\\l\neq i}}^N\frac{dX_i}{dt_1}\frac{dX_l}{dt_1}e^{\Psi^i+\Psi^l}\bigg[U'(X_i-X_l)-U'(X_l-X_i)+\big(U(X_i-X_l)+U(X_l-X_i)\big)\sum_{\substack{j=1\\j\neq i}}^NW'(X_i-X_j)\bigg] \nonumber
            \\&+\sum_{\substack{l\neq k\\l,k\neq i}}\frac{dX_l}{dt_1}\frac{dX_k}{dt_1}e^{\Psi^l+\Psi^k}U(X_l-X_k)\bigg[W'(X_l-X_i)+W'(X_k-X_i)\bigg]\;,\label{Hgf_Pdot_2}
        \end{align}
        where $\Psi^i = \sum_{j=1\;|\;j\neq i}^N\Big(W(X_i-X_j)-V(X_i-X_j)\Big)$. The equation of motion for $t_1$-component will can be obtained by applying the $t_1$-derivative to \eqref{Hgf_Xdot_1}
        \begin{equation}
            \frac{d^2X_i}{dt_1^2} = \frac{dX_i}{dt_1}\sum_{\substack{j=1\\j\neq i}}^N\frac{dX_j}{dt_1}\bigg(V'(X_j-X_i)-V'(X_i-X_j)\bigg)\;.\label{Hgf_EoM1}
        \end{equation}
        We then figure out that the equation \eqref{Hgf_Xdot_2} is nothing but the constraint. The equation of motion for $t_2$-component can be given by performing the derivative in $t_2$ on \eqref{Hgf_Xdot_2}, resulting in
        \begin{align}
            \frac{\partial^2X_i}{\partial t_2\partial t_1} =& 2\frac{dX_i}{dt_1}\sum_{\substack{j\neq l\\j,l\neq i}}^N\frac{dX_j}{dt_1}\frac{dX_l}{dt_1}e^{\Psi^i+\Psi^j}\bigg[U(X_i-X_j)+U(X_j-X_i)\bigg]\bigg[V'(X_l-X_i)-W'(X_i-X_l) \nonumber
            \\&\;\;\;\;\;+V'(X_l-X_j)-W'(X_j-X_l)\bigg]\nonumber 
            \\&+\left(\frac{dX_i}{dt_1}\right)^2\sum_{\substack{j=1\\j\neq i}}^N\frac{dX_j}{dt_1}e^{\Psi^i+\Psi^j}\big[U(X_i-X_j)+U(X_j-X_i)\big] \nonumber
            \\&\;\;\;\;\;\times\left[\frac{\partial\Psi^i}{\partial X_i}+V'(X_i-X_j)-W'(X_j-X_i)+\frac{U'(X_i-X_j)-U'(X_j-X_i)}{U(X_i-X_j)+U(X_j-X_i)}\right] \nonumber
            \\&+\frac{dX_i}{dt_1}\sum_{\substack{j=1\\j\neq i}}^N\left(\frac{dX_j}{dt_1}\right)^2e^{\Psi^i+\Psi^j}\big[U(X_i-X_j)+U(X_j-X_i)\big] \nonumber
            \\&\;\;\;\;\;\times\left[\frac{\partial\Psi^i}{\partial X_i}+V'(X_j-X_i)-W'(X_i-X_j)-\frac{U'(X_i-X_j)-U'(X_j-X_i)}{U(X_i-X_j)+U(X_j-X_i)}\right]\;.\label{Hgf2_EoM2}
        \end{align}
        which is nothing but the equation of motion in $t_3$-dynamics. 
        The above will be used to solve potentials further.
        \\
        \\
        \\
        \\
        \\
        \textbf{The RS system}
        \\
        The generalised Hamilton equations provide
        \begin{align}
            \frac{dX_i}{dt_1} =&\; e^{P_i+\sum_{\substack{j=1\\j\neq i}}^NV_\lambda(X_i-X_j)}\;,\label{Hrs_Xdot_1}
            \\\frac{dX_i}{dt_2} =& \frac{2}{\lambda}\left(\frac{dX_i}{dt_1}\right)^2e^{2\Psi^i_\lambda}+\frac{dX_i}{dt_1}\sum_{\substack{j=1\\j\neq i}}\frac{dX_j}{dt_1}e^{\Psi^i_\lambda+\Psi^j_\lambda}\big[U_\lambda(X_i-X_j)+U_\lambda(X_j-X_i)\big]\;,\label{Hrs_Xdot_2}
            \\\frac{dP_i}{dt_1} =& -\frac{dX_i}{dt_1}\sum_{\substack{j=1\\j\neq i}}^NV'_\lambda(X_i-X_j)+\sum_{\substack{j=1\\j\neq i}}^N\frac{dX_j}{dt_1}\;V'_\lambda(X_j-X_i)\;,\label{Hrs_Pdot_1}
            \\\frac{dP_i}{dt_2} =& -\frac{dX_i}{dt_2}\sum_{\substack{j=1\\j\neq i}}^NW'_\lambda(X_i-X_j) + \sum_{\substack{j=1\\j\neq i}}^N\frac{dX_j}{dt_2}W'_\lambda(X_j-X_i) \nonumber
            \\&+ \sum_{\substack{j=1\\j\neq i}}\sum_{\substack{l=1\\l\neq j\neq i}}\frac{dX_j}{dt_1}\frac{dX_l}{dt_1}e^{\Psi^j_\lambda+\Psi_\lambda^l}\big[U_\lambda(X_j-X_l)W'_\lambda(X_l-X_i)-U_\lambda(X_l-X_j)W'_\lambda(X_j-X_i)\big]\;,\label{Hrs_Pdot_2}
        \end{align}
        where $\Psi^i_\lambda = \sum_{j=1\;|\;j\neq i}\Big(W_\lambda(X_i-X_j)-V_\lambda(X_i-X_j)\Big)$ and the equation \eqref{Hrs_Xdot_2} can be considered as the constraint.
        Taking $t_1$-derivative on \eqref{Hrs_Xdot_1}, we obtain the equation of motion for $t_1$-component
        \begin{equation}
            \frac{d^2X_i}{dt_1^2} = \frac{dX_i}{dt_1}\sum_{\substack{j=1\\j\neq i}}^N\frac{dX_j}{dt_1}\bigg(V'_\lambda(X_j-X_i)-V'_\lambda(X_i-X_j)\bigg)\;.\label{Hrs_EoM1}
        \end{equation}
        And, performing $t_2$-derivative on \eqref{Hrs_Xdot_2}, we obtain
        \begin{align}
            \frac{\partial^2X_i}{\partial t_2\partial t_1} =& \frac{4}{\lambda}\left(2\frac{\partial X_i}{\partial t_1}\frac{d^2X_i}{dt_1^2}e^{2\Psi^i_\lambda}+\sum_{\substack{j=1\\j\neq i}}^N\left(\frac{\partial X_i}{\partial t_1}-\frac{\partial X_j}{\partial t_1}\right)\left(\frac{\partial X_i}{\partial t_1}\right)^2e^{2\Psi^i_\lambda}\frac{\partial \Psi^i_\lambda}{\partial X_i}\right) \nonumber
            \\&+2\frac{dX_i}{dt_1}\sum_{\substack{j\neq l\\j,l\neq i}}^N\frac{dX_j}{dt_1}\frac{dX_l}{dt_1}e^{\Psi^i_\lambda+\Psi^j_\lambda}\bigg[U_\lambda(X_i-X_j)+U_\lambda(X_j-X_i)\bigg]\bigg[V'_\lambda(X_l-X_i)-W'_\lambda(X_i-X_l) \nonumber
            \\&\;\;\;\;\;+V'_\lambda(X_l-X_j)-W'_\lambda(X_j-X_l)\bigg]\nonumber 
            \\&+\left(\frac{dX_i}{dt_1}\right)^2\sum_{\substack{j=1\\j\neq i}}^N\frac{dX_j}{dt_1}e^{\Psi^i_\lambda+\Psi^j_\lambda}\big[U_\lambda(X_i-X_j)+U_\lambda(X_j-X_i)\big] \nonumber
            \\&\;\;\;\;\;\times\left[\frac{\partial\Psi^i_\lambda}{\partial X_i}+V'_\lambda(X_i-X_j)-W'_\lambda(X_j-X_i)+\frac{U'_\lambda(X_i-X_j)-U'_\lambda(X_j-X_i)}{U_\lambda(X_i-X_j)+U_\lambda(X_j-X_i)}\right] \nonumber
            \\&+\frac{dX_i}{dt_1}\sum_{\substack{j=1\\j\neq i}}^N\left(\frac{dX_j}{dt_1}\right)^2e^{\Psi^i_\lambda+\Psi^j_\lambda}\big[U_\lambda(X_i-X_j)+U_\lambda(X_j-X_i)\big] \nonumber
            \\&\;\;\;\;\;\times\left[\frac{\partial\Psi^i_\lambda}{\partial X_i}+V'_\lambda(X_j-X_i)-W'_\lambda(X_i-X_j)-\frac{U'_\lambda(X_i-X_j)-U'_\lambda(X_j-X_i)}{U_\lambda(X_i-X_j)+U_\lambda(X_j-X_i)}\right]\;.\label{Hrs_EoM2}
        \end{align}
        which is the equation of motion for $t_2$-component.
         \\
        \\
        $\bullet$ Now, with all provided relations and Hamiltonian commuting flows, we are going to derive potential functions as follows:
        \\
        \textbf{The CM system}
        \\        
        By employing the consistency, \eqref{cm_pdot3} and \eqref{Hcm_EoM3_2} provide
        \begin{equation}
            W(x)=\frac{3}{2}V(x) + \text{constant}\;.
        \end{equation}
        Now, we consider the Hamiltonian commuting flows \eqref{Hcommute},
           $$\left\{H_1,H_2\right\} = \sum_{i=1}^N\left(-\frac{\partial P_k}{\partial t_2}\frac{\partial X_k}{\partial t_3}+\frac{\partial P_k}{\partial t_3}\frac{\partial X_k}{\partial t_2}\right) = 0\;,$$
        with an \textit{on-shell} condition, i.e., that is substituted by \eqref{cm_xdot2}-\eqref{cm_xdot3}. Thus, we obtain \eqref{Lcm_cyclicell} and, therefore, the elliptic potential \eqref{Lcm_sol}. Thus, we find that what we obtain here, within the Hamiltonian perspective, a perfect match with the Lagrangian one.
         %
        \\
        \\
        \textbf{The GF system}
        \\
        We first consider the consistency (which is equivalent to the Hamiltonian commuting flows \eqref{Hcommute} $\{H_1,H_2\} = 0\;$ with \textit{on-shell} condition, i.e., substituted by \eqref{Hgf_Xdot_1}-\eqref{Hgf_Pdot_2}) resulting in
        \begin{eqnarray}
            0 &=&\sum_{1\le i<l\le N}A(X_i-X_l)\left(\frac{dX_i}{dt_1}\right)^2\frac{dX_l}{dt_1}+\sum_{1\le l<i\le N}B(X_i-X_l)\frac{dX_i}{dt_1}\left(\frac{dX_l}{dt_1}\right)^2\nonumber\\
            &&+\sum_{\substack{i\neq l\neq k}}^NC(X_i-X_l,X_l-X_k)\frac{dX_i}{dt_1}\frac{dX_l}{dt_1}\frac{dX_k}{dt_1}\;,\label{Hgf_condi}
        \end{eqnarray}
        where
        \begin{subequations}
            \begin{align}
                \bar A(x) =& e^{\Psi^i+\Psi^l}\left\{\big[U(x)+U(-x)\big]\left(-\frac{\partial\Psi^i}{\partial X_i}-V'(x)\right)-U'(x)+U'(-x)\right\}\;,\\
                \bar B(x) =& e^{\Psi^i+\Psi^l}\left\{\big[U(x)+U(-x)\big]\left(-\frac{\partial\Psi^l}{\partial X_l}-V'(-x)\right)+U'(x)-U'(-x)\right\}\;,\\
                \bar C(x,y) =& e^{\Psi^i+\Psi^l}\bigg\{\Big(U(x)+U(-x)\Big)\Big(W'(z)+W'(y)-V'(-z)-V'(-y)\Big) \nonumber
                \\&+\Big(U(y)+U(-y)\Big)\Big(W'(-z)+W'(-x)-V'(z)-V'(x)\Big) \nonumber
                \\&+\Big(U(z)+U(-z)\Big)\Big(W'(x)+W'(-y)-V'(-x)-V'(y)\Big)\bigg\}\;.\label{Cbar}
            \end{align}
        \end{subequations}
        We insert \eqref{Lgf_cor} into the conditions $A = 0$ and $B = 0$, in particular $N=3$, providing a bit different relation to \eqref{LgfVUrela} such that
        \begin{equation}
            \Phi(x) := \frac{1}{2}\Big(V(x)+V(-x)\Big) = -\ln\Big(U(x)+U(-x)\Big)+\text{constant}\;.\label{Hgf_phi2}
        \end{equation}
        We now plug \eqref{Hgf_phi2} into the condition $C=0$, see \eqref{Cbar}. It is now not difficult to see that the integrable potential \eqref{Lgf_sol} is obtained.
        \\
        \\
        Furthermore, we consider the Hamiltonian commuting flows \eqref{Hcommute} $\{H_1,H_2\}=0\;$, plugged in by \eqref{Hgf_Xdot_1}-\eqref{Hgf_Pdot_2}. Again, we obtain \eqref{Hgf_condi} and, hence, the elliptic potential \eqref{Lgf_sol}. Thus, the consistency of the Hamiltonian commuting flows and generalised Hamilton equations leads to the selection of elliptic integrability. At this stage, we summarise that, from the Hamiltonian perspective, the elliptic potential is identical to that of the Lagrangian one.
        \\
        \\
        \\
        \\
        \textbf{The RS system}
        \\
        By requiring their consistency, we obtain the relation (which again is equivalent to the Hamiltonian commuting flows \eqref{Hcommute} $\{H_1,H_2\} = 0\;$ substituted by \eqref{Hrs_Xdot_1}-\eqref{Hrs_Pdot_2}) as
        \begin{align}
            0 =& \sum_{i=1}^NA_\lambda(X_i)\frac{2}{\lambda}\left(\frac{dX_i}{dt_1}\right)^3+\sum_{i\neq j}^NB_\lambda(X_i-X_j)\frac{2}{\lambda}\left(\frac{dX_i}{dt_1}\right)^2\frac{dX_j}{dt_1}+\sum_{1\le i<l\le N}C_\lambda(X_i-X_l)\left(\frac{dX_i}{dt_1}\right)^2\frac{dX_l}{dt_1} \nonumber
            \\&+\sum_{1\le l<i\le N}D_\lambda(X_i-X_l)\frac{dX_i}{dt_1}\left(\frac{dX_l}{dt_1}\right)^2+\sum_{\substack{i\neq l\neq k}}^NE_\lambda(X_i-X_l,X_l-X_k)\frac{dX_i}{dt_1}\frac{dX_l}{dt_1}\frac{dX_k}{dt_1}\;,\label{Hrs_condi}
        \end{align}
        where (with $N=3$)
        \begin{subequations}
            \begin{align}
               &\bar A_\lambda(X_i) = e^{\Psi_\lambda^i}\sum_{\substack{j=1\\j\neq i}}^3\big[V'_\lambda(X_i-X_j)-W'_\lambda(X_i-X_j)\big]\;,\\
              &\bar B_\lambda(x) = e^{2\Psi_\lambda^i}\big[-V'_\lambda(x)-W'_\lambda(-x)\big]\;,\\
            &\bar C_\lambda(x) = e^{\Psi_\lambda^i+\Psi_\lambda^l}\left\{\big[U_\lambda(x)+U_\lambda(-x)\big]\left(-\frac{\partial\Psi_\lambda^i}{\partial X_i}-V'_\lambda(x)\right)-U'_\lambda(x)+U'_\lambda(-x)\right\}\;,\\
                &\bar D_\lambda(x) = e^{\Psi_\lambda^i+\Psi_\lambda^l}\left\{\big[U_\lambda(x)+U_\lambda(-x)\big]\left(-\frac{\partial\Psi_\lambda^l}{\partial X_l}-V'_\lambda(-x)\right)+U'_\lambda(x)-U'_\lambda(-x)\right\}\;,\\
               &\bar E_\lambda(x,y) = e^{\Psi_\lambda^i+\Psi_\lambda^l}\bigg\{\Big(U_\lambda(x)+U_\lambda(-x)\Big)\Big(W'_\lambda(z)+W'_\lambda(y)-V'_\lambda(-z)-V'_\lambda(-y)\Big) \nonumber
                \\&\;\;\;\;\;\;\;\;\;\;\;\;+\Big(U_\lambda(y)+U_\lambda(-y)\Big)\Big(W'_\lambda(-z)+W'_\lambda(-x)-V'_\lambda(z)-V'_\lambda(x)\Big) \nonumber
                \\&\;\;\;\;\;\;\;\;\;\;\;\;+\Big(U_\lambda(z)+U_\lambda(-z)\Big)\Big(W'_\lambda(x)+W'_\lambda(-y)-V'_\lambda(-x)-V'_\lambda(y)\Big)\bigg\}\;.
            \end{align}
        \end{subequations}
        The conditions $\bar A_\lambda=0$ and $\bar B_\lambda=0$ would give
        \begin{equation}
            \sum_{\substack{j=1\\j\neq i}}^3\Big(V'_\lambda(X_i-X_j)-W'_\lambda(X_i-X_j)\Big) = 0\;,
        \end{equation}
        and
        \begin{equation}
            V'_\lambda(X_j-X_i)+W'_\lambda(X_i-X_j) = 0\;,
        \end{equation}
        respectively. Therefore, we have 
        $
            V'_\lambda(x) = V'_\lambda(-x)\;
        $
        and
        $
            W_\lambda(x) = V_\lambda(x)+\text{const.}\;
        $
        Next, the conditions $\bar C_\lambda=0$ and $\bar D_\lambda=0$ provide the functional relation of potentials:
        \begin{equation}
            \Phi_\lambda(x) := \frac{1}{2}\Big(V_\lambda(x)+V_\lambda(-x)\Big) = -\ln\Big(U_\lambda(x)+U_\lambda(-x)\Big) + \text{constant}\;,
        \end{equation}
        which is slightly different to \eqref{LrsVUrela}. And, by the condition $\bar E_\lambda=0$, we obtain again the functional relation \eqref{Lrs_func} and, therefore, integrable potential \eqref{Lrs_sol} is obtained. 
        \\
        \\
        To summarise, the \textit{ansatz} of the potential terms in the first two Hamiltonians in the hierarchy together with basic postulates of the variational principle of the Hamiltonian 1-forms immediately lead to elliptic integrable potentials for the CM, GF, and RS systems.
	\section{Concluding discussion}\label{section4}
        In the present paper, the concrete examples of $N$-body in one dimension with long-range 
        interaction, e.g., CM, GF, and RS systems, are studied from the viewpoint of 
        the Langragian multiform formalism, in order to derive the integrable 
        cases from a general ansatz. 
        The generalised Euler-Lagrange (gEL) equations arising from the 
        Lagrangian 1-form structure, in 
        each of these models, leads to a set of compatible equations from which 
        the integrable potentials can be derived. The CM system case is the simplest, 
        but in this case we find that the gEL only provide a weak form of the 
        condition on the potentials, derived from higher order compatibility of the 
        system. In this case the closure relation is needed to specify the 
        integrable potentials in strong form. In the other two cases, the GF mddel 
        and the RS model, the analysis is a bit more involved, but the closure relation is not needed to derive the integrable potentials. 
        \\
        \\
        With regard to the Hamiltonian framework, the integrable elliptic potential is obtained by applying the set of compatible equations, i.e., the generalised Hamilton equations and the Hamiltonian commuting flows, provided in section \ref{ssection3.1}. Additionally, this section also illustrates the Legendre transformation leading to the 2 first Hamiltonians in the hierarchy. The \textit{ansatz} potentials in the CM model can be figured out as the elliptic form with the compatible equations. In equivalence to the Lagrangian context, the Hamilton equations yield the relation of potentials inserting to commuting flows, holding on equations of motion, to obtain elliptic potentials. In the GF model, the \textit{ansatz} of the Hamiltonians comes from \cite{doi:10.2991/jnmp.2005.12.s1.49, Jairuk1}.  The same procedure as in the 
        Lagrangian framework can be used to derive the integrable elliptic potentials. One advantage of the Hamiltonian approach is that the constraint comes directly from the Hamilton equations, while the commutativity of 
        the Hamiltonian flows arises from their Poisson involutivity, leading to 
        the same results as derived from the Lagrangian approach.

        %
	
	\appendix
        \section{The derivation of generalised Euler-Lagrange equation}\label{Appendix}
        In this appendix, we will give some details regarding the derivation of the gEL equations \eqref{contEL}, see also \cite{B_Suris_2013}. Considering the criticality condition 
        on the action functional 
        \eqref{action}, written in parametrised form (parametrising the curve $\Gamma$), and taking the variation $\mathbf{X}\mapsto\mathbf{X}+\delta\mathbf{X}$, we obtain up 
        to first order in the variations:     
        \begin{equation}
            \delta S = 0 = \int_0^1ds \sum_{i=1}^N\left(\frac{\partial L_i}{\partial \mathbf{X}}\frac{dt_i}{ds}\delta\mathbf{X}+\sum_{j=1}^N\frac{\partial L_i}{\partial(\partial\mathbf{X}/\partial t_j)}\frac{dt_i}{ds}\frac{\partial(\delta\mathbf{X})}{\partial t_j}\right)\;.\label{varX1}
        \end{equation}
        Consider the relation
        \begin{equation}
            \frac{d(\delta\mathbf{X})}{ds}\frac{d t_j}{ds} = \left\|\frac{d\mathbf{t}}{ds}\right\|^2\frac{\partial(\delta\mathbf{X})}{\partial t_j}+\sum_{k=1}^N\frac{dt_k}{ds}\left(\frac{dt_j}{ds}\frac{\partial(\delta\mathbf{X})}{\partial t_k}-\frac{dt_k}{ds}\frac{\partial(\delta\mathbf{X})}{\partial t_j}\right)\;,
        \end{equation}
        then \eqref{varX1} becomes
        \begin{equation}
            0 = \int_0^1ds \sum_{i=1}^N\left\{\frac{\partial L_i}{\partial \mathbf{X}}\frac{dt_i}{ds}\delta\mathbf{X}+\frac{1}{\|d\mathbf{t}/ds\|^2}\sum_{j=1}^N\frac{\partial L_i}{\partial(\partial\mathbf{X}/\partial t_j)}\frac{dt_i}{ds}\left(\left(\frac{d(\delta\mathbf{X})}{ds}\right)\frac{dt_j}{ds}-\sum_{k=1}^N\frac{dt_k}{ds}\delta \mathbf{y}_{jk}\right)\right\}\;,\label{varXY}
        \end{equation}
        where
        $$\delta\mathbf{y}_{jk} = -\delta\mathbf{y}_{kj} = \frac{dt_j}{ds}\frac{\partial(\delta\mathbf{X})}{\partial t_k}-\frac{dt_k}{ds}\frac{\partial(\delta\mathbf{X})}{\partial t_j}\;.$$
        Integrating by parts the first term inside the sum within the bracket of \eqref{varXY} and employing the boundary conditions $\delta\mathbf{X}(s(0)) = \delta\mathbf{X}(s(1)) = 0$, the coefficient of $\delta\mathbf{X}$ gives
        \begin{align}
         0 = \sum_{i=1}^N\frac{\partial L_i}{\partial \mathbf{X}}\frac{dt_i}{ds}-\frac{d}{ds}\left\{\frac{1}{\|d\mathbf{t}/ds\|^2}\sum_{i,j=1}^N\frac{\partial L_i}{\partial(\partial\mathbf{X}/\partial t_j)}\left(\frac{d t_i}{ds}\right)\left(\frac{d t_j}{ds}\right)\right\}\;,\label{GEL}
         \end{align}    
        whereas the coefficient of $\delta\mathbf{y}_{jk}$ gives 
    \begin{align}
         0 =& \frac{\partial L_k}{\partial(\partial\mathbf{X}/\partial t_j)}\left(\frac{d t_k}{ds}\right)^2 - \left(\frac{\partial L_k}{\partial(\partial\mathbf{X}/\partial t_k)}-\frac{\partial L_j}{\partial(\partial\mathbf{X}/\partial t_j)}\right)\frac{d t_k}{ds}\frac{d t_j}{ds}  - \frac{\partial L_j}{\partial(\partial\mathbf{X}/\partial t_k)}\left(\frac{d t_j}{ds}\right)^2 \nonumber
         \\&+ \sum_{\substack{i=1\\ i\neq j, i\neq k}}^N\left(\frac{\partial L_i}{\partial(\partial\mathbf{X}/\partial t_j)}\frac{d t_k}{ds}-\frac{\partial L_i}{\partial(\partial\mathbf{X}/\partial t_k)}\frac{d t_j}{ds}\right)\frac{dt_i}{ds}\;.\label{CT}
        \end{align}  
        Equations \eqref{GEL} and \eqref{CT}, respectively, are the gEL equations and the constraint equations.
	\section*{Acknowledgements}
	T. Kongkoom is supported and has received funding support from the NSRF via the Program Management Unit for Human Resources and Institutional Development, Research and Innovation [grant number B37G 660013]. FWN was supported by the Foreign Expert Program of the Ministry of Sciences and Technology of China, grant number G2023013065L.  
	\\
	\\
	Conflict of Interest: The authors declare that they have no
	conflicts of interest.

	\bibliographystyle{unsrt}
	\bibliography{bibliography.bib}

\begin{thebibliography}{10}

\bibitem{S.Lobb}
S.~Lobb and F.~W. Nijhoff.
\newblock {L}agrangian multiforms and multidimensional consistency.
\newblock {\em Journal of Physics A: Mathematical and Theoretical}, 42(45):454013, 2009.

\bibitem{Lobb_2009}
S.~Lobb, F.~W. Nijhoff, and G.~R.~W. Quispel.
\newblock {L}agrangian multiform structure for the lattice {KP} system.
\newblock {\em Journal of Physics A: Mathematical and Theoretical}, 42(47):472002, 2009.

\bibitem{RinCM}
S.~Yoo-Kong, S.~Lobb, and F.~W. Nijhoff.
\newblock {D}iscrete-time {C}alogero-{M}oser system and {L}agrangian 1-form structure.
\newblock {\em Journal of Physics A: Mathematical and Theoretical}, 44(36):365203, 2011.

\bibitem{Rinthesis}
S.~Yoo-Kong.
\newblock {\em {C}alogero-{M}oser type systems, associated {KP} systems, and {L}agrangian structures}.
\newblock PhD thesis, University of Leeds, 2016.

\bibitem{RinRS}
S.~Yoo-Kong and F.~W. Nijhoff.
\newblock Discrete-time ruijsenaars-schneider system and lagrangian 1-form structure.
\newblock {\em arXiv: Exactly Solvable and Integrable Systems}, 2011.

\bibitem{Krich}
I.~Krichever.
\newblock {E}lliptic solutions of the {K}adomtsev-{P}etviashvili equation and integrable systems of particles.
\newblock {\em Functional Analysis and Its Applications}, 14:282--290, 1980.

\bibitem{Xenitidis}
P.~Xenitidis, F.~W. Nijhoff, and S.~Lobb.
\newblock {O}n the {L}agrangian formulation of multidimensionally consistent systems.
\newblock {\em Proceedings of the Royal Society A: Mathematical, Physical and Engineering Sciences}, 467(2135):3295--3317, 2011.

\bibitem{Suris1}
Y.~B. Suris.
\newblock {V}ariational formulation of commuting {H}amiltonian flows: multi-time {L}agrangian 1-forms.
\newblock {\em Journal of Geometric Mechanics}, 5(3):365--379, 2013.

\bibitem{Jairuk2}
U.~Jairuk, M.~Tanasittikosol, and S.~Yoo-Kong.
\newblock {O}n the {L}agrangian 1-form structure of the hyperbolic {C}alogero-moser system.
\newblock {\em Reports on Mathematical Physics}, 79(3):299--330, 2017.

\bibitem{Jairuk3}
U.~Jairuk.
\newblock {\em {L}agrangian 1-form structure for {C}alogero-{M}oser type system}.
\newblock PhD thesis, King Mongkut’s University of Technology Thonburi, 2016.

\bibitem{Petrera}
M.~Petrera and Y.~B. Suris.
\newblock {V}ariational symmetries and pluri-{L}agrangian systems in classical mechanics.
\newblock {\em Journal of Nonlinear Mathematical Physics}, 24(Supplement 1):121--145, 2021.

\bibitem{Suris2}
Yuri~B. Suris.
\newblock Variational symmetries and pluri-lagrangian systems.
\newblock {\em arXiv: Mathematical Physics}, 2013.

\bibitem{Mats2}
M.~Vermeeren.
\newblock {A} variational perspective on continuum limits of {ABS} and lattice {GD} equations.
\newblock {\em Symmetry, Integrability and Geometry: Methods and Applications}, 15:044, 2019.

\bibitem{Mats3}
M.~Vermeeren.
\newblock {H}amiltonian structures for integrable hierarchies of lagrangian {PDEs}.
\newblock {\em Open Communications in Nonlinear Mathematical Physics}, 1:7491, 2021.

\bibitem{10.1093/integr/xyy020}
M.~Vermeeren.
\newblock {C}ontinuum limits of pluri-lagrangian systems.
\newblock {\em Journal of Integrable Systems}, 4(1):xyy020, 2019.

\bibitem{Suris_2018}
Y.~B. Suris.
\newblock {D}iscrete time {T}oda systems.
\newblock {\em Journal of Physics A: Mathematical and Theoretical}, 51(33):333001, 2018.

\bibitem{Jairuk1}
U.~Jairuk, S.~Yoo-Kong, and M.~Tanasittikosol.
\newblock {T}he {L}agrangian structure of {C}alogero's goldfish model.
\newblock {\em Theoretical and Mathematical Physics}, 183(2):665--683, 2015.

\bibitem{PIENSUK202145}
W.~Piensuk and S.~Yoo-Kong.
\newblock {G}eodesic compatibility: {G}oldfish systems.
\newblock {\em Reports on Mathematical Physics}, 87(1):45--58, 2021.

\bibitem{Arnoldtextbook}
K.~Vogtmann, A.~Weinstein, and V.~I. Arnol'd.
\newblock {\em {M}athematical methods of classical mechanics}.
\newblock Graduate Texts in Mathematics. Springer New York, 2010.

\bibitem{Puttarprom2019IntegrableHH}
C.~Puttarprom, W.~Piensuk, and S.~Yoo-Kong.
\newblock {I}ntegrable {H}amiltonian hierarchies and {L}agrangian 1-forms.
\newblock {\em arXiv: Mathematical Physics}, 2019.

\bibitem{Frankbook}
J.~Hietarinta, N.~Joshi, and F.~W. Nijhoff.
\newblock {\em {D}iscrete systems and integrability}.
\newblock Cambridge University Press, 2016.

\bibitem{Suris_2016}
Y.~B. Suris and M.~Vermeeren.
\newblock {\em {O}n the {L}agrangian structure of integrable hierarchies}.
\newblock Springer Berlin Heidelberg, 2016.

\bibitem{Dickey}
L.A. Dickey.
\newblock {\em {S}oliton equations and {H}amiltonian systems}.
\newblock Advanced series in mathematical physics. World Scientific, 2003.

\bibitem{Caudrelier_2020}
V.~Caudrelier and M.~Stoppato.
\newblock {H}amiltonian multiform description of an integrable hierarchy.
\newblock {\em Journal of Mathematical Physics}, 61(12):123506, 2020.

\bibitem{caudrelier2024lagrangian}
V.~Caudrelier, M.~Dell'Atti, and A.~A. Singh.
\newblock {L}agrangian multiforms on coadjoint orbits for finite-dimensional integrable systems.
\newblock {\em Letters in Mathematical Physics}, 114(34):1--52, 2024.

\bibitem{Sleigh_2019_2}
Du. Sleigh, F.~W. Nijhoff, and V.~Caudrelier.
\newblock {V}ariational symmetries and {L}agrangian multiforms.
\newblock {\em Letters in Mathematical Physics}, 110(4):805–826, 2019.

\bibitem{Caudrelier_2024}
V.~Caudrelier, M.~Stoppato, and B.~Vicedo.
\newblock {C}lassical {Y}ang–{B}axter equation, {L}agrangian multiforms and ultralocal integrable hierarchies.
\newblock {\em Communications in Mathematical Physics}, 405(1):12, 2024.

\bibitem{doi:10.2991/jnmp.2005.12.s1.49}
Y.~B. Suris.
\newblock {T}ime discretization of {F}. {C}alogero’s “{G}oldfish” system.
\newblock {\em Journal of Nonlinear Mathematical Physics}, 12(Supplement 1):633--647, 2005.

\bibitem{B_Suris_2013}
Y.~B. Suris.
\newblock {V}ariational formulation of commuting hamiltonian flows: multi-time {L}agrangian 1-forms.
\newblock {\em Journal of Geometric Mechanics}, 5(3):365–379, 2013.

\end{thebibliography}

\end{document}